\documentclass[a4paper,fleqn,usenatbib]{mnras}
\usepackage{newtxtext,newtxmath,hyperref}
\usepackage[T1]{fontenc}
\usepackage{ae,aecompl}
\usepackage[dvipdfmx]{graphicx}	
\usepackage{amsmath}	
\usepackage{amssymb}	
\usepackage{multirow,cases,colortbl}
\usepackage{url}
\usepackage{bm}
\usepackage{lscape}

\usepackage{ulem,color}

\newcommand{\gr}{\ensuremath{\gamma}-ray }
\newcommand{\grs}{\ensuremath{\gamma}-rays }
\newcommand{\fermi}{{\textit{Fermi}}}

\newcommand{\thej}{\ensuremath{{\theta}_{\rm{j}}}}
\newcommand{\theo}{\ensuremath{{\theta}_{\rm{obs}}}}
\newcommand{\thec}{\ensuremath{{\theta}_{\rm{c}}}}

\title[sGRBs similar to GRB 170817A]{On short GRBs similar to GRB 170817A detected by Fermi GBM}

\author[Matsumoto \& Piran]{Tatsuya Matsumoto$^{1,2,3,4}$\thanks{E-mail: tatsuya.matsumoto@mail.huji.ac.il} and Tsvi Piran$^{1}$
\\
$^{1}$Racah Institute of Physics, Hebrew University, Jerusalem, 91904, Israel\\
$^{2}$Research Center for the Early Universe, Graduate School of Science, University of Tokyo, Tokyo 113-0033, Japan\\
$^{3}$Department of Physics, Graduate School of Science, University of Tokyo, Tokyo 113-0033, Japan\\
$^{4}$JSPS Research Fellow\\
}
\pubyear{2019}

\begin{document}
\label{firstpage}
\pagerange{\pageref{firstpage}--\pageref{lastpage}}
\maketitle

\begin{abstract}
\cite{Kienlin+2019} selected 11 short gamma-ray bursts (sGRBs) whose characteristics are similar to GRB 170817A. These bursts, like GRB 170817A, have a hard spike followed by a soft thermal tail. However, as their redshifts  are unknown  it is not clear if their luminosities are as low as that of GRB 170817A. Comparing the positions in the $\epsilon_{\rm p}$-$E_{\rm\gamma,iso}$ (spectral peak energy - isotropic-equivalent  energy) plane and using compactness arguments to estimate the minimal Lorentz factor, $\Gamma$, we find that all the bursts in this sample are consistent with being regular sGRBs if they are located at $z\simeq0.3-3$. They are also consistent with being  similar to  GRB 170817A if they are located at  $z\lesssim0.1$. Even in the latter case, the events must involve at least mildly relativistic ($\Gamma \gtrsim 2$) motion within the sources. 
We, further, find that at most one or two bursts in the sample are consistent with the cocoon shock-breakout model.
Finally, we calculate the event rate of either off-axis emission from a jet core or from a  jet-wing (surrounding the core). We find that the off-axis emission model as an origin of the sample is rejected as it predicts too small event rate. The wing model can be consistent with the observed rate but the model parameters cannot be constrained by the current observations.
\end{abstract}


\section{Introduction}
The first detected gravitational waves (GWs) from a binary neutron star (NS) merger, GW170817 \citep{Abbott+2017c} was accompanied by a \gr counterpart, short gamma-ray burst (sGRB) 170817A \citep{Abbott+2017e,Goldstein+2017,Savchenko+2017}.
Although binary NS mergers were suggested to be a progenitor of sGRBs many years ago \citep{Eichler+1989}, the detected \gr signal was very different from regular sGRBs.
The isotropic-equivalent \gr energy is about three orders-of-magnitude smaller than the weakest event.
This burst also shows an unusual spectral evolution, a hard spike followed by a soft thermal tail  (see below).

Several theoretical models have been  proposed to explain the peculiarity of GRB 170817A.
Focusing on the small energetics, some authors considered an off-axis emission scenario, where we detected the \grs at the outside of the beaming cone of a regular sGRB jet \citep{Goldstein+2017,Murguia-Berthier+2017b,Ioka&Nakamura2018}. However, this model  is ruled out by compactness considerations \citep{Kasliwal+2017,Matsumoto+2019,Matsumoto+2019b}.
Other leading model is the cocoon shock-breakout scenario \citep{Kasliwal+2017,Bromberg+2018,Gottlieb+2018b,Kathirgamaraju+2018,Lazzati+2018,Nakar+2018,Pozanenko+2018}, where the \grs are emitted when a cocoon, produced via the interaction of the jet with ejecta, breaks out from fast expanding ejecta.
In addition to the small total \gr energy, this model can naturally explain the spectral evolution of the burst.
Both of these observational and theoretical studies strongly suggest that the \grs were emitted not from a jet core but from other component in the outflow.

The detection of the unusual GRB 170817A invoked searches for similar events among the archival databases of previously detected GRBs.
\cite{Burns+2018,Troja+2018b} reported that GRB 150101B has a similar \gr spectral evolution to GRB 170817A, although the energetics of this burst is not as small as GRB 170817A.
\cite{Matsumoto+2019b} suggest that the observed $\gamma$-rays in this burst have also been produced via a shock breakout.
Recently, \cite{Kienlin+2019} searched  the 10 yr \textit{Fermi}-gamma-ray burst monitor (\fermi-GBM) burst catalog\footnote{\url{https://heasarc.gsfc.nasa.gov/W3Browse/fermi/fermigbrst.html}}  for  events similar to GRB 170817A.
After several selection processes, they finally picked up 11 events which have a similar \gr spectral evolution to GRB 170817A.
Hereafter we denote these bursts as the vK19 sample.

The important property delineating the vK19 sample and GRBs 150101B and 170817A from other regular sGRBs is their spectral evolution.
These events show  a short and hard non-thermal spike followed by a  longer soft tail whose spectrum is consistent with a thermal spectrum.
This hard-to-soft spectral evolution,\footnote{Commonly, a hard-to-soft spectral evolution \citep[e.g.,][]{Ford+1995,Kaneko+2006,Lu+2012} is referred to   a decaying temporal behavior of the spectral peak energy within a fixed spectral model such as Band or Comptonized spectrum. This is milder  and different than the spectral evolution observed in  GRB 150101B and 170817A, in which the spectral shape  varies from Comptonized to thermal one and the peak energy drops sharply.} so-called the (soft) spectral lag, is a remarkable signature and not common among regular sGRBs  \citep{Bernardini+2015}.
Although some regular sGRBs show soft spectral lags, their evolution is not so fast and dramatic as the one observed in the vK19 sample and GRBs 150101B and 170817A \citep{Burns+2018}.
Some sGRBs show a thermal component, but most of those are coeval with the main non-thermal one \citep[e.g.,][]{Guiriec+2013}.

In this paper, we study the similarity of the vK19 sample and GRB 170817A and explore also the possibility that these events are regular sGRBs.
We organize the paper as follow.
In \S \ref{sample}, we describe the properties of events selected by \cite{Kienlin+2019}.
We check whether these events are consistent with regular sGRBs by inspecting the location in the $\epsilon_{\rm p}$-$E_{\rm\gamma,iso}$  (spectral peak energy - the isotropic-equivalent \gr energy) plane  and  calculating their minimal Lorentz factors using compactness considerations in \S \ref{regular}.
In \S \ref{cocoon}, we study the consistency of the sample with the cocoon shock-breakout model.
In \S \ref{rate}, we calculate the event rates of the off-axis emission and \gr emission from jet-wings surrounding the core  of a jet. Finally we summarize this work in \S \ref{summary}.

\section{The selected events}\label{sample}
GRB 170817A was detected by the \gr detectors on \textit{Fermi} and \textit{INTEGRAL} $\sim1.7\,\rm s$ after the GWs from a binary NS merger \citep{Abbott+2017e,Goldstein+2017,Savchenko+2017}.
The most striking property of this \gr signal is its unusually low isotropic-equivalent energy of $E_{\gamma,\rm iso}\simeq5\times10^{46}\,\rm erg$, which is about three orders-of-magnitude smaller than the weakest sGRB. 
In addition to the small energetics, the GRB also shows a unique spectral signature.
The light curve is composed of two different parts, a hard spike followed by  soft thermal tail.
Detailed temporal and spectral analyses show that the first pulse has a duration of $\simeq0.5\,\rm s$ and a non-thermal spectrum (single power-law with an exponential cutoff, so-called the Comptonized spectrum) with a spectral peak energy $\epsilon_{\rm p}=185\pm62\,\rm keV$ and a spectral index $\alpha=-0.62\pm0.40$.
The following soft component has a duration of $\simeq1\,\rm s$ and its spectrum is consistent with a blackbody spectrum with a temperature $T=10.3\pm1.5\,\rm keV$.
While the properties of the hard spike are common among other sGRBs detected by \fermi-GBM, the transition from the non-thermal spectrum to the blackbody one is rare.

Motivated by GRB 170817A, several authors searched among the previous events for sGRBs with similar signatures to those of GRB 170817A. 
\cite{Troja+2018b,Burns+2018} noticed that GRB 150101B can be such an event.
Although the \gr energy of GRB 150101B is $\sim10^3$ times larger than that of GRB 170817A, thus the energetics is not so peculiar, it is still at the faint end of the sGRB luminosity function.
More importantly, the spectral signature of GRB 150101B is similar to those of GRB 170817A. 
The burst shows a hard spike with a Comptonized spectrum with $\epsilon_{\rm p}=550\pm190\,\rm keV$ and $\alpha=-0.8\pm0.2$ followed by a soft tail with a blackbody spectrum with $T=6.0\pm0.6\,\rm keV$.
Modeling the afterglow, \cite{Troja+2018b} proposed that this burst is an off-axis emission event as proposed for GRB 170817A just after the detection.  However as mentioned earlier,  the off-axis model for GRB 170817A is already ruled out \citep{Kasliwal+2017,Matsumoto+2019,Matsumoto+2019b}.
\cite{Matsumoto+2019b} have shown that  GRB 150101B,  the off-axis scenario is highly unlikely due to compactness considerations and suggested that like GRB 170817A the \gr have been produced by a cocoon breakout.

\cite{Kienlin+2019} searched 10 yr \fermi-GBM burst catalog  picking up candidates  using the duration $T_{90}$ to be in a similar range to GRB 170817A. 
However, too many events satisfy the criterion, partly because $T_{90}$ is calculated  using only the $50-300$ keV energy band, which represents the property of the main spike. 
Then, they manually selected candidates which satisfy the following criteria:
(1) There is a clear luminous spike, which is brighter in the $50-300$ keV band than in the $8-50$ keV band; (2) A weak tail, which is bright in the $8-50$ keV range,  follows the spike ; (3) There is a discernible spectral change between these parts,  avoiding  GRBs showing a continuous hard-to-soft spectral evolution.
After an additional localization test (the main and soft pulses should be localized to the same sky position) and a spectral test of the soft tail (the tail should be well fitted by a thermal spectrum and the temperature should be less than 20 keV), they finally selected 11 sGRBs among 395 sGRBs detected by \fermi-GBM.
Table \ref{table list} shows the selected events (the vK19 sample) and the properties of their first pulses.
All events have a first spike with a Comptonized spectrum, and do not have a measured redshift.
In the vK19 sample, GRB 170111B shows soft emissions before and after it hard spike. Although this emission episode is different from that of GRB 170817A, we include this event in our analyses for completeness. 

\begin{table*}
\begin{center}
\caption{The events selected by \citealt{Kienlin+2019} (vK19 sample). Observables are only for first hard spikes. }
\label{table list}
\begin{tabular}{lrrrrrr}
\hline
Event&Energy flux&Fluence &Duration &Minimal variable timescale&Peak energy&Spectral index\\
&$F$ [$10^{-7}\,\rm{erg\,cm^{-2}\,s^{-1}}$]&$S$ [$10^{-7}\,\rm{erg\,cm^{-2}}$]&$\delta t$ [s]&$\delta t_{\rm min}$ [ms]&$\epsilon_{\rm p}$ [keV]&$\alpha_p$ \\
\hline\hline
GRB 081209A& $40.14\pm0.20$ &$15.4\pm0.8$& $0.384$ & $<15$ & $1473\pm275$ & $-0.75\pm0.08$\\
GRB 100328A& $33.61\pm1.70$ &$15.1\pm0.8$& $0.448$ & $<11$ & $927\pm177$ & $-0.54\pm0.10$\\
GRB 101224A& $4.04\pm0.15$ &$2.07\pm0.77$& $0.512$ & $47$ & $341\pm320$ & $-1.04\pm0.39$\\
GRB 110717A& $25.57\pm3.50$ &$3.27\pm0.45$& $0.128$ & $11$ & $328\pm67$ & $-0.34\pm0.26$\\
GRB 111024C& $4.11\pm0.53$ &$1.58\pm0.20$& $0.384$ & $41$ & $144\pm18$ & $0.53\pm0.60$\\
GRB 120302B& $3.72\pm0.53$ &$1.90\pm0.27$& $0.512$ & $<120$ & $133\pm20$ & $0.66\pm0.68$\\
GRB 120915A& $13.66\pm1.40$ &$6.99\pm0.72$& $0.512$ & $41$ & $526\pm114$ & $-0.21\pm0.25$\\
GRB 130502A& $2.12\pm0.30$ &$3.26\pm0.46$& $1.536$ & $221$ & $91\pm20$ & $-0.80\pm0.35$\\
GRB 140511A& $14.85\pm1.70$ &$2.85\pm0.33$& $0.192$ & $<94$ &$280\pm58$ & $-0.78\pm0.16$\\
GRB 170111B& $7.42\pm0.70$ &$3.80\pm0.36$& $0.512$ & $<63$ & $154\pm22$ & $-0.62\pm0.19$\\
GRB 180511A& $34.29\pm3.80$ &$2.19\pm0.24$& $0.064$ & $<5$ & $639\pm220$ & $-0.61\pm0.22$\\
\hline
\end{tabular}
\end{center}
\end{table*}

\section{Can the vK19 sample be regular sGRBs?}\label{regular}
\subsection{Location in the $\epsilon_{\rm p}$-$E_{\gamma,\rm iso}$ plot}\label{amati}
\begin{figure}
\begin{center}
\includegraphics[width=85mm]{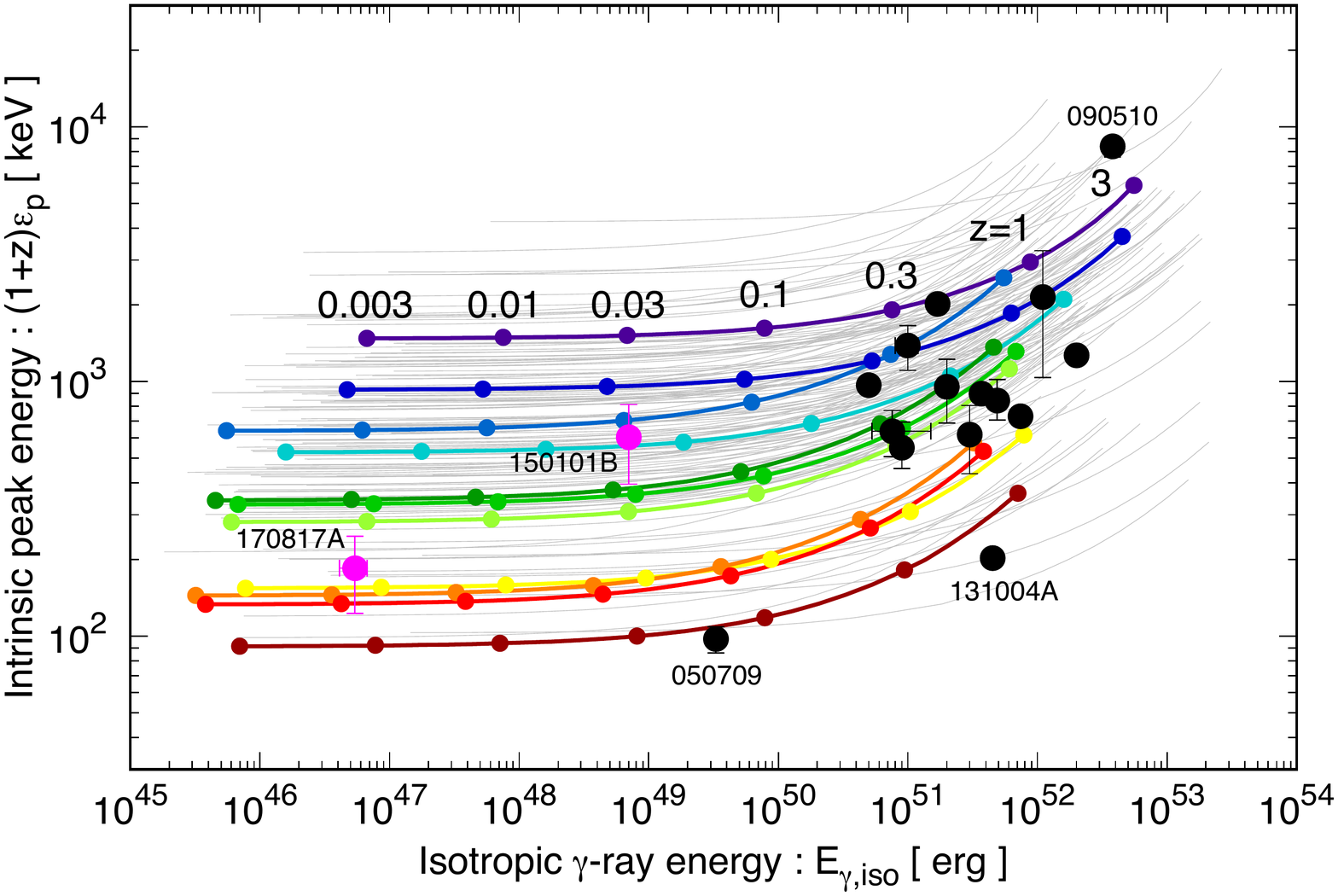}
\caption{Trajectories of bursts in the $\epsilon_{\rm p}$-$E_{\gamma,\rm iso}$ plane.
Each colored trajectory shows the sequence of 11 events in the vK19 sample for redshifts $0.003 \leq z \leq 3$.
The colors correspond to the peak energy (bluer is a higher peak energy) and are the same as in Fig. \ref{fig gamma_b}.
Grey curves show other 143 sGRBs with a Comptonized spectrum detected by \fermi-GBM.
sGRBs with measured redshifts \citep[taken from][]{Zhang+2012,Tsutsui+2013,Fong+2015}  are shown with black dots including GRBs 170817A and 150101B (magenta).
Note that GRB 130603B with the macronova/kilonova candidate \citep{Berger+2013,Tanvir+2013} is among the central  group of sGRBs.
 }
\label{fig amati}
\end{center}
\end{figure}

First, we study the locations of the vK19 events in the $\epsilon_{\rm p}$-$E_{\gamma,\rm iso}$  (spectral peak energy - isotropic-equivalent \gr energy) plane. Fig. \ref{fig amati} depicts the redshift trajectories of the 11 events as well as other sGRBs on this plane when the redshift is varied over the range  $0.003\leq z\leq3$.
We also show in this figure normal sGRBs with measured redshifts denoted by black dots and trajectories of others 
whose redshift is unknown as  grey trajectories.
These normal events are taken from \cite{Zhang+2012,Tsutsui+2013,Fong+2015}, and the 10 yr \fermi-GBM burst catalog \citep{Gruber+2014,Kienlin+2014,Bhat+2016}.
Since the vK19 events have a Comptonized spectrum, we consider only regular  sGRBs with a Comptonized spectrum (143 events).
Most of the  sGRBs with a known redshift seem to be located in a region\footnote{A possible correlation between $\epsilon_{\rm p}(1+z)$ and $E_{\gamma,\rm iso}$ like the Amati relation which is proposed  for long GRBs \citep{Amati+2002}, has been suggested for sGRBs \citep[e.g.,][]{Zhang+2012,Tsutsui+2013}. However it is not obvious in Fig. \ref{fig amati}.} with $\epsilon_{\rm p}(1+z)\sim400-1000\,\rm keV$ and $E_{\gamma,\rm iso}\simeq10^{51-52}\,\rm erg$ while there are some outliers, GRBs 050709, 090510, and 131004A.

The vK19's trajectories are located in a similar position to those of regular ones.
In particular, they overlap the range of  sGRBs with known redshift if their redshifts are $z\simeq0.3-3$ as expected for regular sGRBs.
It should be noted that the peak energy is determined from the first hard pulse of the burst, whose properties are not so peculiar compared with those of normal sGRBs.
Thus, it is natural that the vK19 bursts overlap the locations of regular sGRBs.
At the redshift $z\lesssim0.1$, some of the vK19 events are located in a similar range to GRBs 170817A and 150101B.

\subsection{Compactness}\label{compactness}
Next, we consider the compactness limit: a \gr emitting site should be optically thin to the observed \grs.
There are several opacity sources  \citep{Lithwick&Sari2001}:
(A) The most energetic photon should escape from the source without producing pairs by colliding with other photons. (B) Photons should not be scattered by the pairs created by high-energy photons. (C) Photons  should not be scattered by electrons accompanying the baryons in the outflow.
Since for selected events the maximal photon energy is not reported in the catalog (and would be small for bursts with a Comptonized spectrum), we consider here only limits B and C.

\begin{figure}
\begin{center}
\includegraphics[width=85mm]{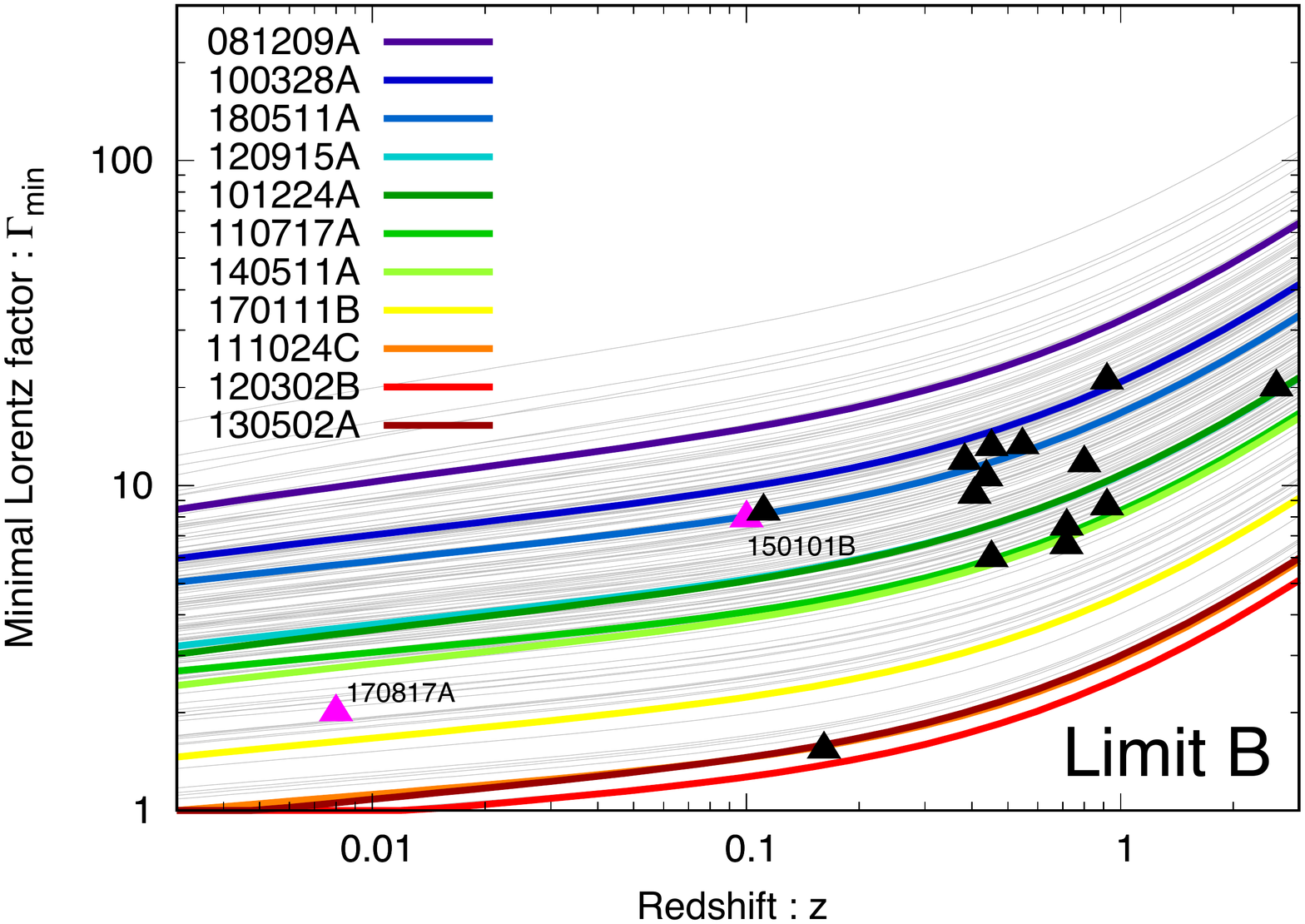}
\includegraphics[width=85mm]{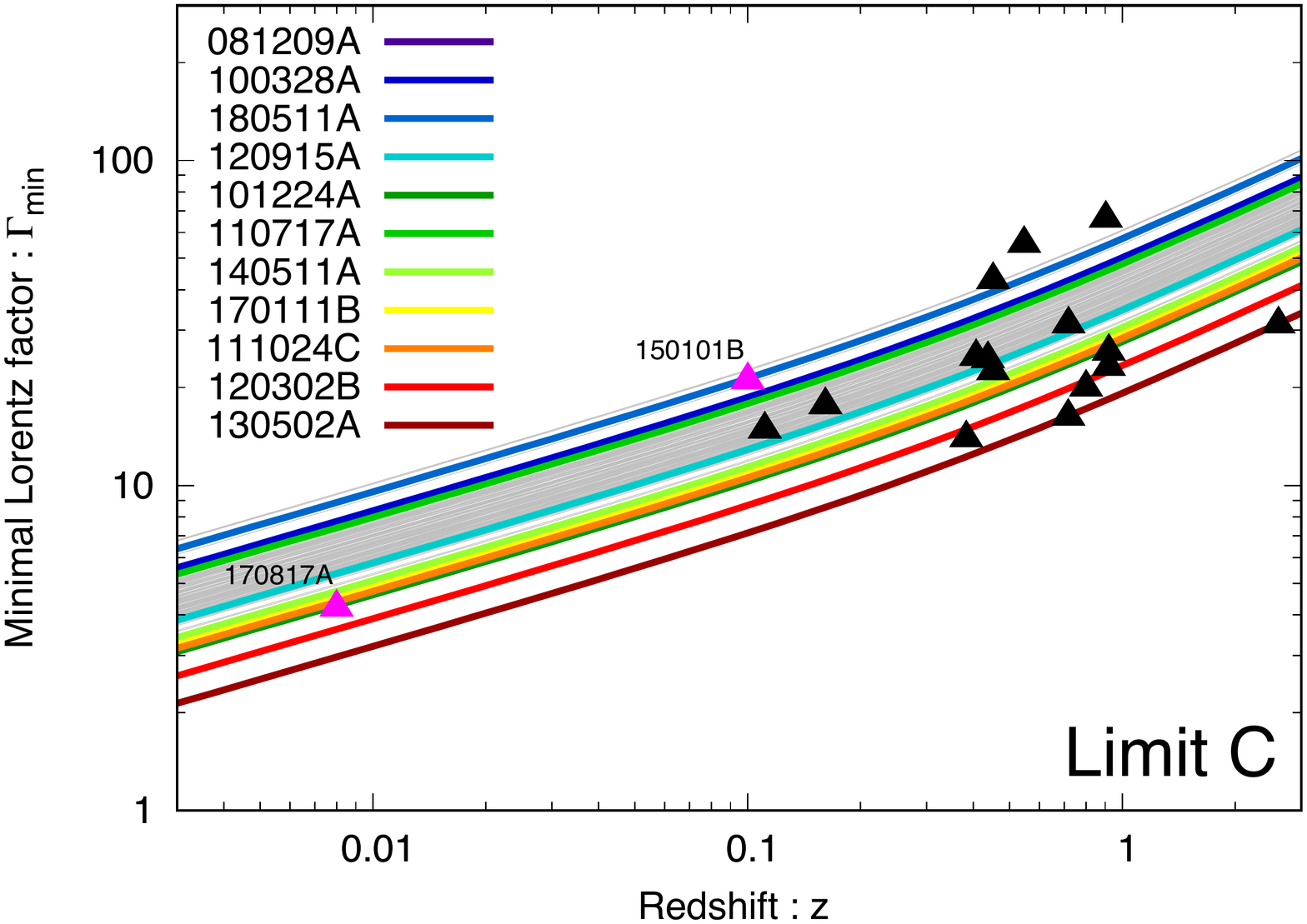}
\caption{{\bf Top}: Minimal Lorentz factor for limit B.
The color solid curves denote the minimal Lorentz factors of the vK19 events as a function of the redshift.
The grey curves also show the minimal Lorentz factors for 143 regular sGRBs without measured redshifts.
Black and magenta triangles denote regular sGRBs with redshifts and GRB 170817A and 150101B, respectively.
{\bf Bottom}: The same as limit C.}
\label{fig gamma_b}
\end{center}
\end{figure}

Fig. \ref{fig gamma_b} depicts the minimal Lorentz factor determined by limits B and C  as a function of the redshift (see \citealt{Matsumoto+2019b} for a detailed discussion).
Generally,  limit C yields the more constraining Lorentz factor.
Only for bursts with a large peak energy,  limit B gives a larger minimal Lorentz factor at small redshift ($z\lesssim0.01$).
As shown in Fig. \ref{fig gamma_b}, even if the redshift is very small, the vK19 events should be relativistic. 
However, the limits on the Lorentz factor are rather modest but this is common for the whole sGRB sample.
We also show the minimal Lorentz factors of other sGRBs with grey curves (without redshift) and black triangles (with redshift).
Clearly, the minimal Lorentz factors of the vK19 sample are consistent with those obtained for regular sGRBs.
Again, this is because the adopted parameters are calculated mainly for first spike, which is not so peculiar compared with normal bursts.

Interestingly, the minimal Lorentz factors of regular sGRBs are smaller than that of long GRBs (typically $\gtrsim100$).
This is already pointed out by \cite{Nakar2007}.
It is attributed to the fact that  the fluences of sGRBs are usually small and their high-energy spectrum is not well determined.
Hence typically it is fitted better by a Comptonized spectrum than by a Band spectrum \citep{Ghirlanda+2004,Ghirlanda+2009}.
Thus this result may come from the observational selection effect, and be changed if more sensitive  \gr detectors in the high-energy range will be available. 
Actually the minimal Lorentz factor of the powerful GRB 090510, which shows a Band spectrum, is constrained to be large $\Gamma_{\rm min}\simeq700$ in limit B (\citealt{Matsumoto+2019b}, see also \citealt{Ackermann+2010} for  other limits).

As a conclusion of the above two tests, all events in the vK19 sample are consistent with being regular sGRBs as shown by their position in the  $\epsilon_{\rm p}$-$E_{\gamma,\rm iso}$ plot (Fig. \ref{fig amati}) and by compactness  considerations (Fig. \ref{fig gamma_b}).
However, they can be also consistent with GRBs 150101B and 170817A, which do not belong to the regular sGRB population (Fig. \ref{fig amati}).

\section{The cocoon shock breakout model}\label{cocoon}
The cocoon shock-breakout model \citep{Kasliwal+2017,Bromberg+2018,Gottlieb+2018b,Kathirgamaraju+2018,Lazzati+2018,Nakar+2018,Pozanenko+2018}.
offers a reasonable explanation for GRB 170817A (see above references) and 150101B \citep{Matsumoto+2019b} and it explains naturally the thermal soft tail of the sample.
Furthermore, in this model, we can estimate the redshift of each event and check the consistency of the scenario in detail.
Thus we turn here to estimate the compatibility of the vK19 sample with this model. 

The closure relation of the relativistic shock-breakout emission \citep{Nakar&Sari2012,Gottlieb+2018b} provides a suitable compatibility test.
The observables of the hard spike should satisfy \citep{Nakar&Sari2012}:
\begin{align}
\delta t^\prime\simeq1{\,\rm s\,}\biggl(\frac{E_{\gamma,\rm iso}}{10^{46}\,\rm erg}\biggl)^{1/2}\biggl(\frac{\epsilon_{\rm p}^\prime}{150\,\rm keV}\biggl)^{-\frac{9+\sqrt{3}}{4}},
   \label{eq closure}
\end{align}
where $\delta t^\prime$ and $\epsilon_{\rm p}^\prime$ are the duration and the spectral peak energy measured in the burst-rest frame, respectively.
These quantities are related to the observed (unprimed) quantities as $\delta t=(1+z)\delta t^\prime$ and $\epsilon_{\rm p}=\epsilon_{\rm p}^\prime/(1+z)$.
The \gr energy is given by $E_{\gamma,\rm iso}=4\pi d_{\rm L}^2Sk(z){/(1+z)}$, where $d_{\rm L}$, $S$, and $k(z)$ are the luminosity distance, the observed fluence, and the $k$-correction for a given spectrum respectively.
The $k$-correction is calculated based on \cite{Bloom+2001}, and it converts the energy in the observed $8-1000\,\rm keV$ energy range to that in the $1-10000\,\rm keV$ energy range in the burst-rest frame.
Substituting them in Eq. \eqref{eq closure} we obtain the following equation for the redshift of the event:
\begin{align}
\biggl(\frac{\delta t}{\rm s}\biggl)^{-1}\biggl(\frac{S}{10^{-7}\,\rm erg\,cm^{-2}}\biggl)^{1/2}&\biggl(\frac{\epsilon_{\rm p}}{150\,\rm keV}\biggl)^{-\frac{9+\sqrt{3}}{4}}k(z)^{1/2}\nonumber\\
&\simeq(1+z)^{\frac{{7}+\sqrt{3}}{4}}\biggl(\frac{d_{\rm L}(z)}{29\,\rm Mpc}\biggl)^{-1}.
   \label{eq closure2}
\end{align}
While the left-hand side depends only on observables ($k$-correction is almost independent of redshift for the spectral parameters of the sample), the right-hand side is a function of the redshift.
The right-hand side has the minimum at $z\simeq1.3$ and there are two redshift values which satisfy this equation.
Here we take the lower one ($z\lesssim1.3$).

\begin{figure}
\begin{center}
\includegraphics[width=85mm]{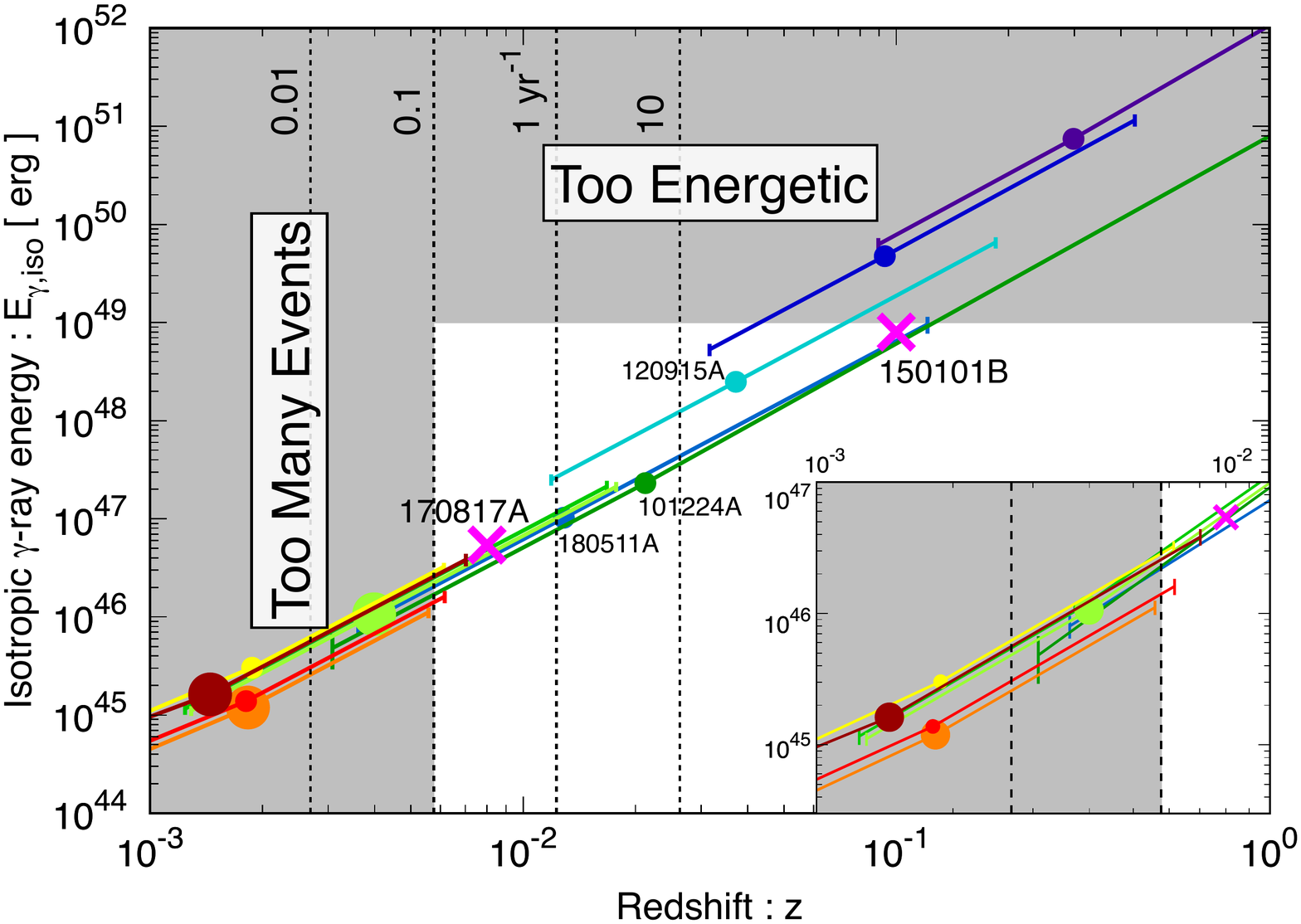}
\caption{Inferred redshift from the closure relation (Eq. \ref{eq closure2}) and isotropic \gr energy of each event.
The vertical lines show the redshifts within which the binary-NS-merger event rates become 0.01, 0.1, 1, and 10$\,\rm yr^{-1}$ for the NS merger rate of $1550\,\rm Gpc^{-3}\,yr^{-1}$.
Clearly, bursts at too small redshift $z\lesssim0.003$ and too energetic bursts with $E_{\gamma,\rm iso}\gtrsim10^{49}\,\rm erg$ are inconsistent with the observed detection rate and the energetics of the cocoon shock-breakout emission.
The bursts shown in small dots are also inconsistent with the cocoon model because of their disconnected light curves or their short time variability.
The inset figure shows the zoom-up of the region of $z\leq10^{-2}$.}
\label{fig cocoon}
\end{center}
\end{figure}

Fig. \ref{fig cocoon} depicts the evaluated redshift and the corresponding $E_{\gamma,\rm iso}$ of the events in the vK19 sample and, for a comparison, those of GRBs 170817A and  150101B.\footnote{The first peak of GRB 150101B is very short and its fluence is dominated by  the first  time-bin (which is about half of the duration of this peak). 
Hence we use its properties: $\epsilon_{\rm p}=1290\pm590\,\rm keV$ and duration of 0.008 s \citep{Matsumoto+2019b}.}
The uncertainty of the redshift determination arises mainly from the uncertainty in the peak energy.
We also take into account an uncertainty by a factor of $\sim2$ in the normalization of the closure relation.
We find that some events are inconsistent with the cocoon shock-breakout model.
First, since only a small fraction of the total kinetic energy is dissipated and emitted in a shock-breakout emission, too energetic events with $E_{\gamma,\rm iso}\gtrsim10^{49}\,\rm erg$ cannot be explained by a shock breakout.
We estimate the critical (maximal) \gr energy as a fraction $10^{-3}$ of typical isotropic-equivalent kinetic energy of sGRB $E_{\rm k,iso}\sim10^{52}\,\rm erg$.
This limit excludes GRBs 081209A and 100328A.

Second, very nearby events with $z\lesssim0.003$ are unreasonable because they require too large binary NS merger rate.
In Fig. \ref{fig cocoon}, we show the redshifts whose corresponding volumes give the merger rates of  0.01, 0.1, 1, and 10 $\rm yr^{-1}$.
We set a fiducial value of the binary NS merger rate as ${\cal R}=1550{\,\rm Gpc^{-3}\,yr^{-1}}$, as follows from the rate obtained by the GW observation as ${\cal R}\simeq1550_{-1220}^{+3220}\,\rm Gpc^{-3}\,yr^{-1}$ \citep{Abbott+2017c}.
For the 10-years \fermi-GBM observation period  only 0.01 event could occur within the redshift $z\lesssim0.003$ while the vK19 sample contains 4 such candidates.
Thus, the events that have too small redshifts, GRBs 111024C, 120302B, 130502A, and 170111B, are excluded within this scenario.
It should be noted that we do not take the beaming effect, viewing angle dependence, or the sky-covering factor of {\fermi}-GBM into account in this estimate.
Since these effects make the GRB event rate smaller than the NS merger rate, the constraints on these 4 events with $z\lesssim0.003$ becomes more severe.
When taken into account, these effects may also lead to the rejection of GRBs 110717A and 140511A ($z\simeq0.004$).
Additionally, the rate of these 6 events is also inconsistent with local sGRB rate \citep[within $200\,\rm Mpc$,][]{Mandhai+2018}.
The rest of the  events with $z\gtrsim0.01$ are consistent with the cocoon shock-breakout scenario.

Within the cocoon shock-breakout model, the hard spike and soft tail are produced in the planer and spherical phases respectively and there is no significant temporal gap between the two. Additionally we do not expect variability on a timescale much shorter than the total duration. 
When considering the light-curve structure, we find that with the nominal values no GRB among the 11 events in the vK19 sample
is consistent with the shock-breakout model.
Some events in the vK19 sample show clearly separated hard spike and soft tail, others show a short time variability \citep{Kienlin+2019} and both types are ruled out.
These considerations rule out GRB 101224A that has a separated light curve and  GRBs 120915A and 180511A  that show a short timescale variability.
When the error-bar of the redshift estimate is taken into account one or two events (GRBs 130502A and 140511A) could be consistent with the shock-breakout scenario.
Thus, the fraction of shock-breakout event rate is less than $\lesssim2/395\simeq0.5\,\%$ of the total \fermi-GBM sGRBs.

\section{Off-axis and  jet-wing event rates}\label{rate}
The ratio of the number of the events in the vK19 sample (and cocoon-emission candidates discussed in the previous section) to the total regular sGRBs is $\sim3\,\%$ ($\lesssim0.5\,\%$).
While the sample is still consistent with regular sGRBs, if the hard spike and following soft tail identified in the vK19 sample are indeed a special signature, they may suggest a novel mechanism for the \gr emission.
Here we consider two possibilities that such a signature results from the geometry of the outflow, and study their
consistency with the event rate of the vK19 sample.
One possibility that has been discussed extensively is that this signature is produced by off-axis emissions, that is, the viewing angle dependence of the event.
Although this possibility is rejected \citep[see e.g.][]{Kasliwal+2017,Matsumoto+2019,Matsumoto+2019b} for GRB 170817A, we check it again for the vK19 sample.
The other is that the bursts are produced by an outflow surrounding the  jet core, hereafter we call this part a jet wing.
The cocoon shock-breakout emission scenario, discussed earlier,  is one such possibility.

We consider the following three simple axisymmetric jet structures and estimate the event rates  for: 
(a) A top-hat jet \citep[see also][]{Salafia+2016}; (b) A Gaussian jet;  (c) A top-hat jet with Gaussian wings (see also e.g., \citealt{Kathirgamaraju+2018} for estimates of the rate of wing emission based on the result of numerical simulations and \citealt{Beniamini+2019} for the joint detection rate of GWs and \grs from NS mergers with various jet structures).
We adopt  simple models in which we assume that the \gr luminosity and Lorentz factor distributions at the source (fluid) rest frame are given by:
(a) A top-hat jet,
\begin{align}
\frac{dL}{d\Omega}&=\begin{cases}
1&;\theta\leq\thej,\\
0&;\text{otherwise,}\\
\end{cases}\\
\Gamma&=\begin{cases}
\Gamma&;\theta\leq\thej,\\
0&;\text{otherwise;}\\
\end{cases}
\end{align}
(b) A Gaussian jet, 
\begin{align}
\frac{dL}{d\Omega}&=\exp\biggl[-\frac{\theta^2}{2{\thej}^2}\biggl],\\
\Gamma&=(\Gamma-1)\exp\biggl[-\frac{\theta^2}{2{\thej}^2}\biggl]+1;
\end{align}
and (c) A top-hat jet with Gaussian wings, 
\begin{align}
\frac{dL}{d\Omega}&=\begin{cases}
1&;\theta\leq\thej,\\
\exp\biggl[-\frac{(\theta-\thej)^2}{2{\thej}^2}\biggl]&;\text{otherwise,}\\
\end{cases}\\
\Gamma&=\begin{cases}
\Gamma&;\theta\leq\thej,\\
(\Gamma-1)\exp\biggl[-\frac{(\theta-\thej)^2}{2{\thej}^2}\biggl]+1&;\text{otherwise.}\\
\end{cases}
\end{align}
Here $\Gamma$ and $\thej$ are the Lorentz factor at the jet axis ($\theta=0$) and the jet opening angle, respectively.
In these simple models, we assume that the luminosity and Lorentz factor distributions have the same function form.
We normalize the luminosity distribution at the jet axis.
Models (a) and (b) are motivated by the off-axis and wing scenarios, respectively.
In model (c), the jet has a comparable width to the wing and core $\sim\thej$ (we give a detailed definition of a jet core below).
This is an intermediate structure between models (a) and (b), and it is useful to understand the result.
We stress that these are ad hoc models and there is no reason that nature will adopt these structure. In particular we consider Gaussians in the lab frame rather than in the source frame. These are motivated by simplicity and the aim of this calculation is to provide an insight to the likelihood of these scenarios.

The observed luminosity at a given viewing angle $\theo$ is obtained by integrating the luminosity distribution taking into account the corresponding Doppler factors as
\begin{align}
L_{\gamma,\rm iso}(\theta_{\rm obs})&=\int d\Omega\,\delta_{\rm D}^4(\Omega,\theta_{\rm obs})\frac{dL}{d\Omega},
   \label{eq lobs}\\
\delta_{\rm D}(\Omega,\theta_{\rm obs})&=\frac{1}{\Gamma(\theta)\big[1-\beta(\theta)\cos\vartheta\big]},
   \label{eq doppler}\\
\cos\vartheta&=\sin\theta_{\rm obs}\sin\theta\cos\phi+\cos\theta_{\rm obs}\cos\theta,
\end{align}
where we use spherical coordinate and locate the observer at ($\theta$, $\phi$)=($\theta_{\rm obs}$, 0).

\begin{figure*}
\begin{center}
\includegraphics[width=185mm]{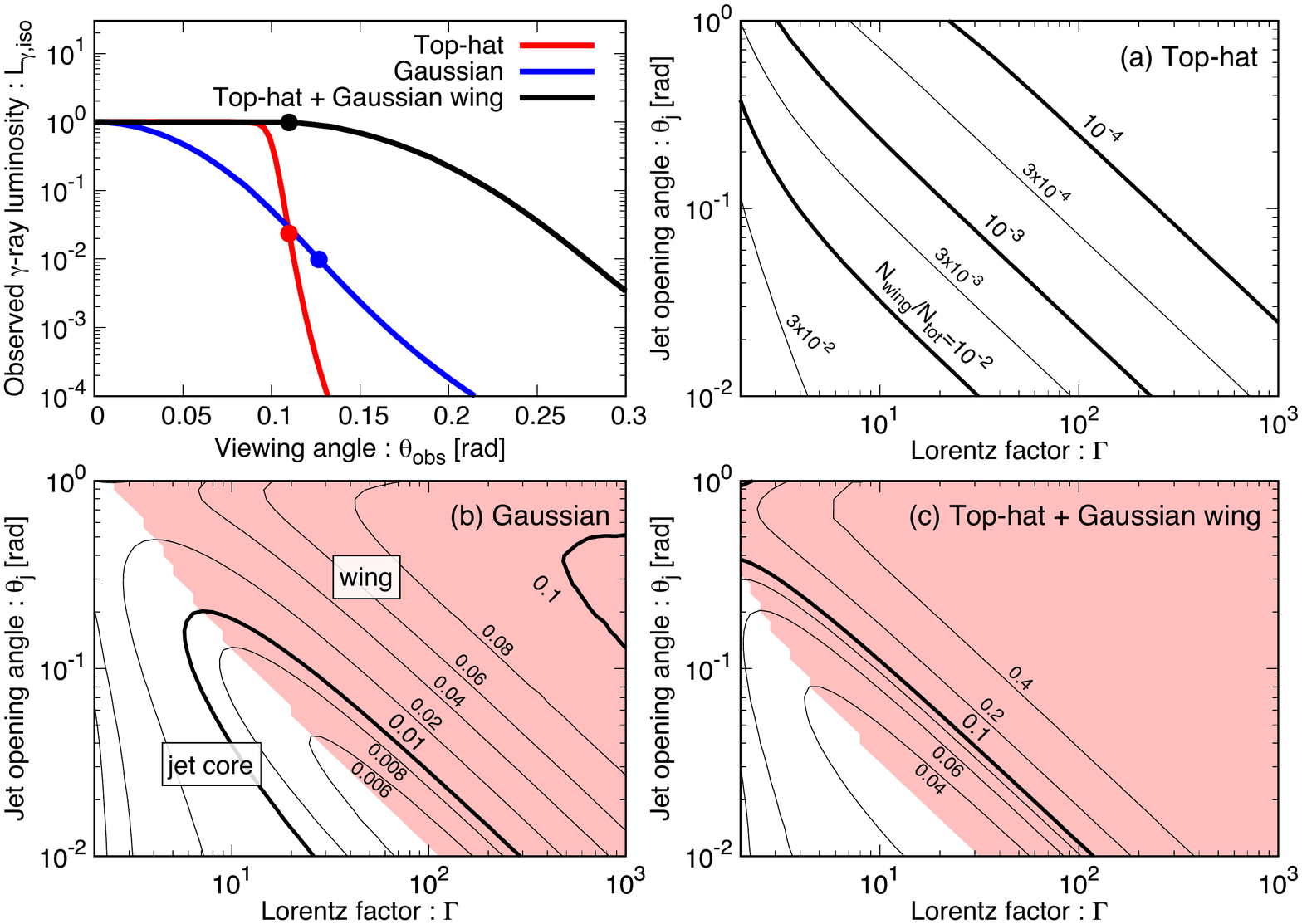}
\caption{{\bf (Top left)} Profiles of observed luminosity $L_{\gamma,\rm iso}$.
The red, black, and blue curves denote the profiles of top-hat, top-hat with Gaussian wing, and Gaussian jets, respectively.
The parameters are set to be $\Gamma=100$ and $\thej=0.1\,\rm rad$.
Each dot shows the position of the jet core $\thec$.
{\bf (Top right)} Contours of fixed event-rate ratio of off-axis (wing) emission to total events $N_{\rm wing}/N_{\rm tot}$ for the top-hat jet model.
{\bf (Bottom)} The same as the top right panel but for the Gaussian jet (left) and the top-hat with a Gaussian wing jet (right) models.
The shaded regions ($\Gamma\thej\gtrsim1$) show a regime where the ratio increases because the emission from ejecta at the line-of-sight (wing) dominates the observed luminosity.}
\label{fig rate}
\end{center}
\end{figure*}

The top left panel of Fig. \ref{fig rate} depicts the profile of the observed luminosity for each jet model with $\thej=0.1\,\rm rad$ and $\Gamma=100$.
The points denote the edge of the observable jet core, which we define as:
\begin{align}
\thec\equiv\thej+1/\Gamma(\thej).
\end{align}
For the top-hat jet, the luminosity declines rapidly beyond the jet core due to the de-beaming effect (we can see a similar behavior for $E_{\gamma,\rm iso}$ in \citealt{Kasliwal+2017,Granot+2017,Ioka&Nakamura2018}).
Since a wing component may dominate the emission outside of the core \citep{Matsumoto+2019,Ioka&Nakamura2019}, the observed luminosity becomes brighter for top-hat with a Gaussian wing jets (black) and for Gaussian jets (blue).

The event rate with for an observer at a viewing angle of $0<\theta_{\rm obs}^\prime<\theo$ is given by 
\begin{align}
N(0,\,\theo)=\int_0^{\theo}d{\theta_{\rm obs}^\prime}\sin\theta_{\rm obs}^\prime{\,\cal R\,}\frac{4\pi}{3}d^3(\theta_{\rm obs}^\prime),
   \label{eq rate}
\end{align}
where $d(\theo)=\sqrt{{L_{\gamma,\rm iso}(\theo)}/{4\pi f}}$ is the maximal distance at which an event with a luminosity $L_{\gamma,\rm iso}(\theo)$ can be detected by a detector with a sensitivity $f$.
Note that the arguments of $N$ are the upper and lower limits of the integral.
In this equation, we have integrated over the distance (or volume) without taking any redshift effects into account \citep[see][for a calculation taking them into account]{Salafia+2016}.
Since a \gr emission produced by a wing may be so weak that it is detectable only at a low redshift, this is a reasonable approximation.
If the binary merger rate $\cal R$ is an increasing function of the redshift, the event rate of the jet-core emissions $N(0,\thec)$ increases accordingly.

We calculate the ratio of the event rate of wing emissions to that of total events, $N_{\rm wing}/N_{\rm tot}$ for each jet model (thus the ratio is independent of the merger rate $\cal R$, sensitivity $f$, and normalization of luminosity distribution).
Here the wing-emission and total event rates are defined by using Eq. \eqref{eq rate} as: 
\begin{align}
N_{\rm wing}&\equiv N(\thec,\pi/2),\\
N_{\rm tot}&\equiv N(0,\pi/2),
\end{align}
respectively.
As shown in this definition of $N_{\rm wing}$, it corresponds the event rate of the emissions observed from the outside of the jet core.
Thus, it  equals to the off-axis emission rate for the top-hat jet model (a).
Clearly, the ratio depends on the definition of jet core $\thec$.
In particular, since $L_{\gamma,\rm iso}$ sharply decreases with the viewing angle,  $N_{\rm wing}$ is dominated by the observed luminosity just outside the jet core $\sim\thec$ .
However, it is uncertain (and unlikely) that a small difference of the viewing angle results in a different emission signature such as an appearance of a soft tail.
We expect that an observed emission near (but just outside of) the jet core will be similar to that from the jet core, and only gradually changes as the viewing angle.
Thus the ratio $N_{\rm wing}/N_{\rm tot}$ would give an upper limit of the ratio of wing or off-axis emission (with soft tails) to total events.

Top right and bottom panels of Fig. \ref{fig rate} depict contours of fixed values of $N_{\rm wing}/N_{\rm tot}$ in the $\Gamma$-$\thej$ plane for different jet models.
For the top-hat jet model (top right panel), which corresponds to the off-axis hypothesis, the contours are roughly given by $\Gamma\thej\simeq\rm const$ and larger $\Gamma$ and $\thej$ give a smaller ratio.
This is understood as follows.
Since the jet core always dominates the observed luminosity, the total event rate is evaluated by the product of the solid angle and the luminosity of the jet core as $N_{\rm tot}\sim\theta_{\rm j}^2$, where we assume the jet-core luminosity as unity.
The wing rate is similarly given as $N_{\rm wing}\sim(\thej/\Gamma) L_{\gamma,\rm iso}^{3/2}(\thec)$, where we approximate the solid angle of the jet wing as a ring with a radius $\thej$ and width of $1/\Gamma$.
The observed \gr luminosity at $\thec$ hardly depends on $\thej$ and $\Gamma$.
Therefore, the ratio is estimated by $N_{\rm wing}/N_{\rm tot}\propto1/(\thej\Gamma)$.

For the Gaussian jet and the top-hat jet with Gaussian wings models (bottom panels in Fig. \ref{fig rate}), the contour maps have a valley at $\Gamma\thej\sim1$.
When  $\Gamma\thej<1$, we can regard the jet as a point source and the observed luminosity is dominated by the jet core.
Thus the observed-luminosity profile $L_{\gamma,\rm iso}$ becomes similar to that of the top-hat jet, which gives the ratio $N_{\rm wing}/N_{\rm tot}\propto1/(\thej\Gamma)$.
On the other hand, when $\Gamma\thej>1$, only the emission from the line-of-sight region mainly contributes to the observed luminosity and $L_{\gamma,\rm iso}$ has a broad wing (see top left panel of Fig. \ref{fig rate}).
Moreover, the edge of the jet core converges to the jet opening angle $\thec\to\thej$ for larger Lorentz factor, which significantly increases $L_{\gamma,\rm iso}(\thec)$ and also boosts $N_{\rm wing}$.
As a consequence, the fraction increases for larger $\Gamma\thej$.
This regime is shown by the pink shaded region in the bottom panels in Fig. \ref{fig rate}. 

From the top-right panel of Fig. \ref{fig rate}, we can clearly reject the off-axis possibility that off-axis emissions produce the vK19 events.
For a reasonable parameter range of $\Gamma\thej\gtrsim1$ (such as $\Gamma=100$ and $\thej=0.1\,\rm rad$), the ratio of rates is much smaller than that of the vK19 sample $\sim3\,\%$.
On the other hand, jet models with wings would be consistent with the observed event-rate ratio because these models give a ratio larger than the observed one.

\section{Summary}\label{summary}
We studied the 11 sGRBs in the vK19 sample and compared them both to 170817A and to regular sGRBs. 
Inspection of the position of these events in the $\epsilon_{\rm p}$-$E_{\gamma,\rm iso}$ plane and compactness considerations show that the vK19 events are consistent with the regular-sGRB population if they are at $z\simeq0.3-3$.
About half of the events are  also consistent with being in the same parameter region as GRB 170817A if they are located at $z\lesssim0.1$ (see Figs. \ref{fig amati} and \ref{fig gamma_b}) and the other half is closer to the parameter region of GRB150101B.
The compactness argument reveals that even if the redshifts of these bursts are very small ($z\lesssim0.01$), the \gr emitting region should be (at least mildly) relativistic, $\Gamma_{\rm min}\gtrsim 2$. Interestingly, given the weak limit on the high energy end of these bursts and of most sGRBs even if at higher redshifts compactness argument gives only modest limits (typically $\Gamma_{\rm min} < 100$ and in cases much lower but always relativistic) on the Lorentz factor of the emitting regions in sGRBs \citep[see][]{Nakar2007}. 
  
The cocoon emission scenario reasonably explains the emission properties of GRB 170817A \citep{Kasliwal+2017,Bromberg+2018,Gottlieb+2018b,Kathirgamaraju+2018,Lazzati+2018,Nakar+2018,Pozanenko+2018}.
This specific scenario provides us with a closure relation of a relativistic shock-breakout emission \citep{Nakar&Sari2012} that enables us to check further the consistency of this model. 
Using this closure relation, we evaluate the redshifts and $E_{\gamma,\rm iso}$ of the vK19 events within this scenario.
Among the 11 events in the sample, 6 bursts imply too low $E_{\gamma,\rm iso}$ and in turn close distances and hence too large detection rate (compared with the event rate implied by GW 170817A and local sGRBs).
Another 2 bursts imply too large $E_{\gamma,\rm iso}$ values that cannot be sustained by this model.
Within the remaining 3 bursts, one event has a too wide gap between the first non-thermal peak and the second thermal peak.
The other two events show a short time variability, which cannot be produced by a shock-breakout.
While there is no obvious candidate of a cocoon shock-breakout emission in the vK19 sample, 
when  the error of the redshift estimate is taken into account,  one or two events may be consistent with this scenario. This is consistent with the possible event rate of nearby binary NS mergers.

We have also confirmed that off-axis emission model in which the observed \grs are emitted from a jet core viewed not along the direction of motion of the jet (namely outside of the beaming cone), results in an extremely small event rate and cannot reproduce the observed event rate of 11 events within the 10 years \fermi-GBM observation. 
On the other hand, a wing emission scenario, in which the jet core has wide wings that also emit \grs, can give a consistent event rate with the observed one.
In this case, given binary NS mergers with the event rate of ${\cal R}\simeq1000\,\rm yr^{-1}\,Gpc^{-3}$, the detection rate of the vK19 sample by {\fermi} $\sim1\,\rm yr^{-1}$ constrain the redshift to be larger than $z\gtrsim0.012$ or $d_{\rm L}\gtrsim54\,\rm Mpc$. Thus the luminosity of these events should larger than $L_{\gamma,\rm iso}\gtrsim5\times10^{46-47}{\,\rm erg\,s^{-1}}$.
However, since the parameter space describing such a model is very large, it is difficult to constrain the luminosity or Lorentz factor distributions from the observed ratio of event rates.

To conclude while the vK19 events can be associated with a wing emission, two of them could be a cocoon shock-breakout events and all can also be simple regular sGRBs. It seems that without a redshift determination it will be impossible to determine the real origin of these events.

\section*{acknowledgments}
We thank Kunal P. Mooley for helpful comments and the anonymous referee for giving us important comments.
TM thank Chi-Ho Chan for useful comments to improve TM's computer skills.
This work is supported by Grant-in-Aid for JSPS Research Fellow 19J00214 (TM) and an advanced ERC grant TReX and by the CHE-Icore center for excellence in Astrophysics (TP).

\bibliographystyle{mnras}
\bibliography{reference_matsumoto}

\bsp	
\label{lastpage}
\end{document}